\documentclass[twocolumn,preprintnumbers,amsmath,amssymb,superscriptaddress]{revtex4}

\usepackage{graphicx}
\usepackage{bm}
\usepackage{color}
\usepackage{ulem}

\begin{document}

\title{Universal nonlinear stage of the locally induced modulational instability in fiber optics}

\author{Adrien E. Kraych}
\affiliation{Univ. Lille, CNRS, UMR 8523 - PhLAM -
 Physique des Lasers Atomes et Mol\'ecules, F-59000 Lille, France}
\author{Pierre Suret}
\affiliation{Univ. Lille, CNRS, UMR 8523 - PhLAM -
 Physique des Lasers Atomes et Mol\'ecules, F-59000 Lille, France}
\author{Gennady El}
\affiliation{Department of Mathematical Sciences, Loughborough
 University, Loughborough LE11 3TU, United Kingdom}
\author{St\'ephane Randoux}
 \email{stephane.randoux@univ-lille1.fr}
\affiliation{Univ. Lille, CNRS, UMR 8523 - PhLAM -
 Physique des Lasers Atomes et Mol\'ecules, F-59000 Lille, France}

\date{\today}

\begin{abstract}
We report an optical fiber experiment in which we study  nonlinear
stage of modulational instability of a plane wave  in the presence of
a localized  perturbation.  Using a recirculating fiber loop as
experimental platform, we show that the initial perturbation evolves
into expanding nonlinear oscillatory structure exhibiting some
universal  characteristics that agree with theoretical predictions
based on integrability properties of the focusing nonlinear
Schr\"odinger equation. Our experimental results demonstrate
persistence of the universal evolution scenario, even in the presence
of small dissipation and noise in an experimental system that is not
rigorously of an integrable nature.
\end{abstract}


\maketitle

Modulational instability (MI), known as the Benjamin-Feir instability
in water waves, is a ubiquitous phenomenon in focusing nonlinear media
that is manifested in the growth of small, long-wavelength
perturbations of a constant background
\cite{BF:67a,BF:67b,Zagryadskaya:68,Ostrovskii:72,Zakharov:09a,Tai:86,Soljacic:00,Solli:12,Meir:04}.
The linear stage of MI is characterized by an exponential growth of
all the perturbations falling in the region of the Fourier spectrum
below certain cut-off wavenumber \cite{Zakharov:09a}.  This simple
picture ceases  to be valid when the amplitude of the growing
perturbation becomes comparable to the background, i.e.  at the
nonlinear stage of MI.

In the nonlinear regime, MI exhibits a rich spatiotemporal dynamics
that has been recently the subject of  significant interest in several
areas of experimental and theoretical physics
\cite{Erkintalo:11,Kibler:10,Frisquet:13,Akhmediev:11,Biondini:15,GEl:16,Coulibaly:15,Frisquet:15,Kimmoun:16,Mussot:18}.
In this respect, the focusing one-dimensional nonlinear
Schr\"{o}dinger equation (1D-NLSE) plays a prominent role as a
universal mathematical model describing at leading  order wave
phenomena relevant to many fields of nonlinear physics such as
e.g. optics and hydrodynamics \cite{Dudley:14}. A particular scenario
of the MI development strongly depends  on the type of initial
conditions considered. In the majority of the existing analytical,
numerical and experimental studies of MI periodic or random initial
modulations of a constant background have been considered
\cite{Erkintalo:11,Kibler:10,Frisquet:13,Akhmediev:11,Frisquet:15,Kimmoun:16,Mussot:18,Soto:16}. For
these types of initial conditions the nonlinear stage of the MI
development was shown to be dominated by breather-like structures such
as the Akhmediev, Kuznetsov-Ma, Peregrine breathers and their
generalizations.  The role of NLSE breather solutions  has been
extensively discussed in the last years in the context of the
formation of rogue waves \cite{Shrira:10,Onorato:13}. A particular
type of breather solutions of the 1D NLSE, the so-called superregular
solitonic solutions  have been shown in \cite{Zakharov:13,Gelash:14}
to describe the development of a certain type small {\it localized}
perturbations of the plane wave \cite{Kibler:15}.

When a localized (and not necessarily small)
initial perturbation of a plane wave has an arbitrary shape (within a
reasonably broad class), it was recently  shown using the inverse
scattering transform solutions of the 1D-NLSE  that the nonlinear
dynamics of MI is characterized by a ``hyperbolic'' scenario, where a
universal (not depending on the shape of the initial localized
perturbation to leading order) nonlinear oscillatory structure
develops  and expands in time with {\it finite speed}
\cite{Biondini:16a,Biondini:16b}.  In sharp contrast with the
previously mentioned MI scenarios involving the formation of various
breathers, this scenario involves the formation of a symmetric
expanding nonlinear wave structure described by the modulated elliptic
solution of the 1D-NLSE. The modulation provides a gradual transition
from a fundamental soliton resting at the center to small-amplitude
dispersive waves propagating away from the center with linear group
velocity.  This universal modulated elliptic solution of the 1D-NLSE
saturating the MI was first obtained in \cite{GEl:93} in the framework
of the Whitham modulation theory \cite{whitham}.
Importantly, considered  for either left or right spatial domain
(assuming the traditional ``mathematical'' notions of space and time
variables in the 1D-NLSE) this modulation solution was also shown to
describe the development of the 1D-NLSE ``focusing dam break'' problem
\cite{Kamchatnov:97, GEl:16, Jenkins:14}.  This remarkable connection
between two apparently unrelated problems provides an additional
insight into the nonlinear dynamics of MI. 

It has been demonstrated in ref. \cite{Biondini:17} that the
qualitative behaviors found within the integrable NLSE framework are
robust and the considered  nonlinear stage of MI can also be found in
a variety of other wave systems being not necessarily integrable. In
view of the fundamental significance of the 1D-NLSE and its
generalizations, the experimental realization of this universal
scenario of the MI development is of major importance for nonlinear
physics.

In this paper, we report the experimental observations of the
space-time dynamics of a modulationally unstable plane wave modified
by two types of localized, real-valued perturbations: a hump and a
well.  Using a recirculating fiber loop as experimental platform, the
perturbed plane wave is propagated over hundreds of kilometers with
only very small power losses.  The behaviors observed experimentally 
are quantitatively very well described by the 1D-NLSE with a small
linear damping term, and our experimental observations reveal that the
wave structure described by the solution considered in
ref. \cite{GEl:93,Kamchatnov:97,Biondini:16a,Biondini:16b,Biondini:17}
is robust to noise and to the deviations from integrability that are
inherent in any experimental system.

Even though modern single mode fibers (SMFs) represent propagation
media with very small linear losses (typically $\sim 0.2$ dB/km at the
telecommunication wavelength of $1550$ nm), their attenuation cannot
be considered as being fully negligible over propagation distances of
a few kilometers. Many of the optical fiber experiments realized those
last years for the observation of breather solutions of the 1D-NLSE
have encompassed this constraint by using waves with an optical power
of the order of $\sim 1$ Watt.  With this power, the characteristic
nonlinear length typically ranges between $100$ m and $1$ km so that
single-pass propagation experiments reasonably well described by the
1D-NLSE can be performed within propagation lengths between one to
several kilometers. 
In all single pass fiber geometries where Watt-level powers are
required, the generated nonlinear structures have typical durations
falling between $\sim 1$ ps and $\sim 10$ ps. This  requires the use
of fast optical detection devices like optical sampling oscillocopes
or time lenses  \cite{Kibler:10,Xu:17,Walczak:15,Suret:16,Nahri:16}.
Moreover the observation of the space-time dynamics in single-pass optical
fiber experiments often represents a difficult task since it relies on
destructive cut-back techniques \cite{Tikan:17} or alternatively, on
nonlinear digital holography methods \cite{Tsang:03,Tikan:18}.

In our work, we have adopted another strategy by implementing a
recirculating fiber loop that presents the significant advantage  to
provide real-time observation of the space-time dynamics of the
optical wave. Recirculating fiber loops  have been previously used
under many circumstances in the context of  optical fiber
communication \cite{Desurvire:85}, in particular to demonstrate
long-distance transmission of solitons
\cite{Mollenauer:88,Nakazawa:91,Okhrimchuk:01,Golovchenko:97}.  Here
this fiber system is used to provide in a non-destructive way the
real-time stroboscopic view of the ``slow'' evolution of the perturbed
plane wave, round trip after round trip inside a passive fiber ring
cavity.  In our recirculating  fiber loop, the optical power is kept
typically around only $\sim 10$ mW and the propagation distances that
are reached  can be as large as hundreds of kilometers. With this
experimental  approach, all the physically-relevant characteristic
lengths and durations are rescaled  by one or two orders of magnitude,
i.e. the nonlinear length becomes  of the order of $\sim 100$ km and
the typical duration of soliton  structures becomes $\sim 50$ ps. With
such a time scale, the local perturbation  of the plane wave can be
relatively easily engineered by using standard fast  electro-optic
modulators (EOMs). Moreover the detection  part can be ensured by fast
electronic devices like  photodiodes and oscilloscopes. 

\begin{figure}[h]
  \includegraphics[width=0.5\textwidth]{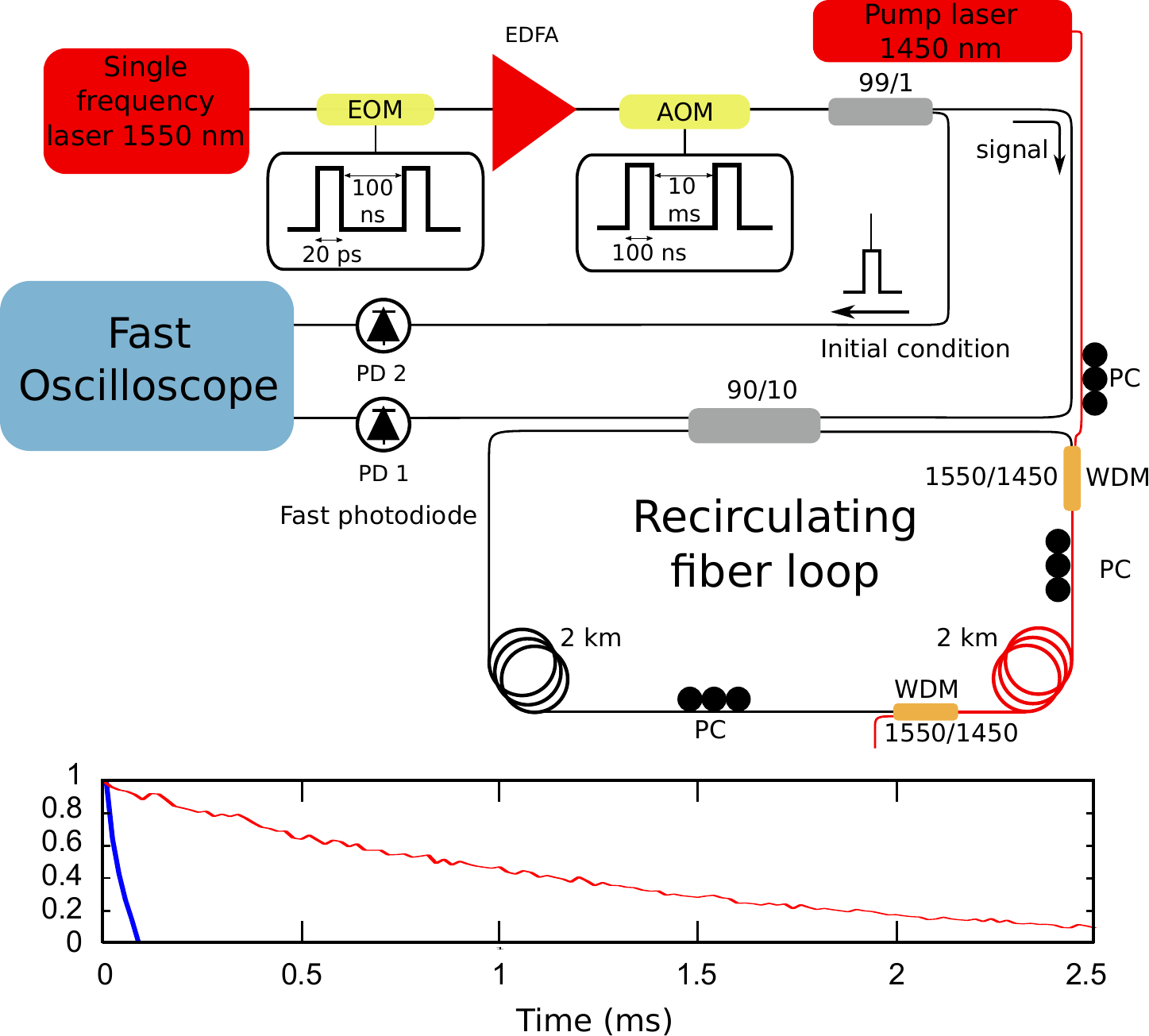}
  \caption{Schematic representation of the experimental setup. PC:
    polarization controller, EDFA: Erbium-doped fiber amplifier, AOM:
    acousto-optic modulator.  The bottom part represents the decay of
    the plane wave  measured in absence (blue line) and in presence
    (red line) of Raman amplification at a pump power of $535$ mW.}
\end{figure}

Our experimental setup is schematically shown in Fig. 1.  It consists
of a recirculating fiber loop, i.e. a passive ring cavity made up of
$\sim 4$ km of SMF closed on itself by a $90/10$ fiber coupler. The
coupler is arranged in such a way that $90 \%$ of the intracavity
power is recirculated. A wide light pulse of $\sim 100$ ns having a
square shape is perturbed by a small localized perturbation of $\sim30$
ps and circulates in the counterclockwise direction inside the
fiber loop. The perturbed square pulse is generated by modulating
the power of a single-frequency laser operating at $1550$ nm, see
Appendix \ref{sec:exp}.

The square pulse plays the role of a pertubated plane wave that is
periodically injected inside the loop with a period  of $10$ ms which
is much larger than the cavity round-trip time of $\sim 20$ $\mu$s. It
is monitored by a fast photodiode (PD1) connected to an oscilloscope
having an electrical bandwidth of $65$ GHz.
This photodiode has been
carefully calibrated and the optical power of the plane wave launched
inside the loop  is known with a relative accuracy that is below
$\sim 10 \%$.

\begin{figure*}[!t]
  \includegraphics[width=1\textwidth]{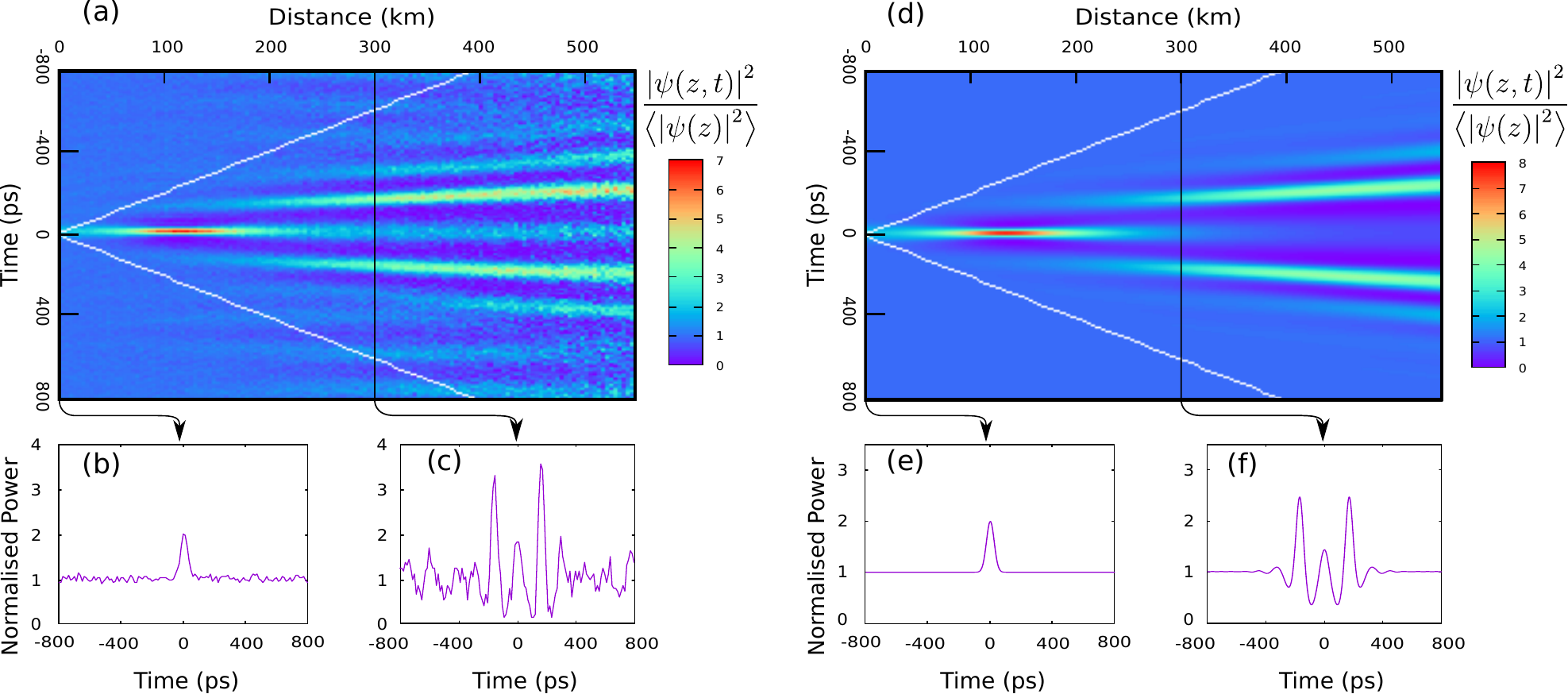}
  \caption{Space-time evolution of a modulationally-unstable plane
    wave perturbed at the initial stage by a localized bright
    (positive) peak.  Left part: (a)(b)(c) experiments and right part:
    (d)(e)(f) numerical simulation of Eq. (\ref{eq:NLSE}) with
    $\beta_2=-22$ ps$^2$km$^{-1}$, $\gamma=1.3$W$^{-1}$km$^{-1}$,
    $\alpha_{eff}=4.2 \times 10^{-3}$ km$^{-1}$, $P_0=14$ mW, $T_0=30$
    ps ($\psi(z=0,t)=\sqrt{P_0 (1 + \exp(-t/T_0)^2)}$).  At each round
    trip inside the cavity in (a) and (d), the optical power has been
    renormalized by the mean  power carried by the
    exponentially-decaying plane wave.
    $<|\psi(z)|^2>=P_0 \exp(-\alpha_{\rm eff} z)$ represents the
    mean power of the plane wave at position $z$.  }
\end{figure*}

The linear losses of the SMF are around $\sim 0.2$ dB/km. They are
partially compensated by Raman amplification in a section of the loop
that is $2-$km long. Following the method used in
ref. \cite{Xu:17,Mussot:18}, a pump beam at $1450$ nm  is launched in
a counterpropagating (clockwise) direction to provide Raman gain with
weak relative intensity noise.  The pump power $P_p$ at $1450$ nm is
typically around $\sim 500$ mW which is much greater than the power
($P_0 \sim 10$ mW) of the plane wave circulating inside the loop.
The pump radiation at $1450$ nm is
coupled in and out the  fiber loop  by using two wavelength
dense multiplexers (WDMs). Increasing the power $P_p$ of the pump beam
at $1450$ nm from zero to a few hundreds of mW, the decay time of the
square pulse
that propagates inside the loop and that is measured by the photodiode
PD2 at the output of the fiber coupler dramatically increases from
$\sim 40$ $\mu$s to $\sim 1$ ms, as shown in the bottom part of
Fig. 1. Let us emphasize that our ring cavity is conceptually
different and also simpler than coherently driven passive cavities
used e. g. in ref. \cite{Leo:10,Copie:16} as intrinsically bistable
devices that can support dissipative cavity solitons.

Fig. 2 (left part) shows the space-time evolution of an optical plane
initially perturbed by a localized bright (positive) perturbation
having a duration of $\sim 30$ ps and a peak power twice as large as
the mean power ($14$ mW) of the plane wave, see Fig. 2(b).  As shown
in Fig. 2(a), the experiment reveals that a nonlinear oscillatory
structure develops from the initial localized perturbation and expands
with propagation distance, in qualitative agreement with the scenario
theoretically described in
refs. \cite{GEl:93,Biondini:16a,Biondini:16b,Biondini:17}.  It should
be noted, however, that the structure observed in the experiment
captures only the initial stage of the development of MI towards the
universal pattern of
\cite{GEl:93,Biondini:16a,Biondini:16b,Biondini:17,Taki:10} with the
fundamental soliton at the center (see Appendix \ref{sec:MI}). Another
comment is that a  localized perturbation of the plane wave realised
in the experiment could generally violate the restriction that the
spectrum of the associated scattering problem must be pure continuous
required for the analysis in \cite{Biondini:16a,Biondini:16b}. With
the above caveats, the observed pattern confirms the robustness of the
universal scenario predicted by the theory.

\begin{figure*}[!t]
  \includegraphics[width=1\textwidth]{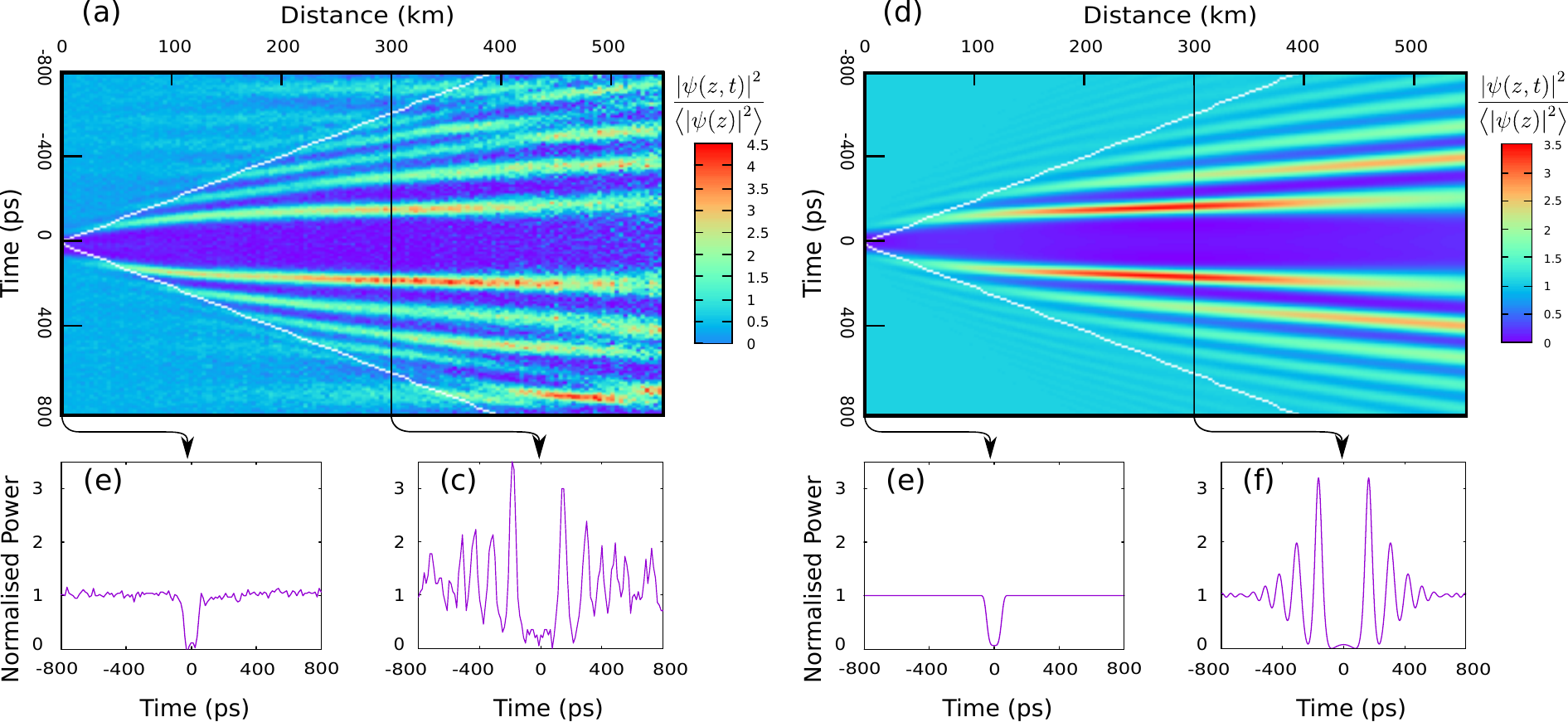}
  \caption{Space-time evolution of a modulationally-unstable plane
    wave perturbed at initial stage by a localized dark (negative)
    peak.  Left part: (a)(b)(c) experiments and right part: (d)(e)(f)
    numerical simulation of Eq. (\ref{eq:NLSE}) with $\beta_2=-22$
    ps$^2$ km$^{-1}$, $\gamma=1.3$W$^{-1}$km$^{-1}$, $\alpha_{eff}=4
    \times 10^{-3}$ km$^{-1}$, $P_0=16$ mW, $T_0=50$ ps, $\beta=0.93$
    ($\psi(z=0,t)=\sqrt{P_0 (1 - \beta \exp(-t/T_0)^4)}$). }
\end{figure*}

As discussed in ref.\cite{Biondini:16b}, the development of the
oscillation behavior of the nonlinear stage of MI under localized
perturbations is significantly influenced by noise. Even a very
small amount of numerical noise inherent in simulations of the 1D-NLSE
with pseudo-spectral methods significantly pertubs the oscillating
structure that is destroyed at finite evolution time
\cite{Biondini:16b}.  In the experiment, the situation is worse in the
sense that {\it (i)} the 1D-NLSE only represents an approximate model
of the experiment and {\it (ii)}, the initial plane wave has a
non-zero noise level. Therefore,  the process of the development of
the nonlinear oscillatory structure within the wedge-shaped region of
Fig. 2(a) can be completely overtaken by the process of the
exponential amplification of the small optical noise that perturbs the
laser field at the initial stage. 

In our experiment, these difficulties have been circumvented by
carefully adjusted the power $P_0$ of the initial plane wave and the
power $P_p$ of the $1450$ nm pump laser.  In Fig. 2(a), $P_0$ and
$P_p$ have been adjusted to $14$ mW and $535$ mW, respectively.  With
these values, the growth rate of the oscillatory structure that
emerges from the local perturbation, the noise amplification rate  and
the cavity loss rate are sufficiently well balanced for the nonlinear
oscillatory structure to be observed over a propagation distance of
$\sim 500$ km.  Any increase of $P_p$ beyond $535$ mW induces some
significant amplification of the noise level which results in the
destruction of the oscillatory structure generated from the nonlinear
evolution of the local perturbation.

Behaviors found in the experiments are retrieved with a good
quantitative agreement from the numerical integration of the 1D-NLSE
with small linear damping:
\begin{equation}\label{eq:NLSE}
  i\frac{\partial \psi}{\partial z}=\frac{\beta_2}{2}\frac{\partial^2
    \psi}{\partial t^2}-\gamma|\psi|^2\psi -i \frac{\alpha_{\rm
      eff}}{2} \psi
\end{equation}
and with parameters corresponding to the experiments. $\psi(z,t)$
represents the complex envelope of the electric field that slowly
varies in space $z$ and time $t$.  At $1550$nm  the  group velocity
dispersion coefficient  of the SMF is $\beta_2=-22$ ps$^2$/km. The
Kerr coefficient  is $\gamma=1.3$W$^{-1}$km$^{-1}$ and the effective
power losses $\alpha_{\rm eff}$ measured from the decay rate of the
plane wave inside the ring cavity are $4.2 \times 10^{-3}$ km$^{-1}$
(equivalently $9.6 \times 10^{-4}$ dB/km).

Fig. 2 (right panel) shows the result of the numerical integration of
Eq. (\ref{eq:NLSE}) by taking $\psi(z=0,t)=\sqrt{P_0 (1 +
  \exp(-t/T_0)^2)}$ as initial condition. Taking $P_0=14$ mW and
$T_0=30$ ps, this expression fits quite well the experimental profile
plotted in Fig. 2(b).  There is a good quantitative agreement between
right (numerical) and left (experimental) parts of Fig. 2, which
indeed confirms that our experiment is well described by
Eq. (\ref{eq:NLSE}) where power losses have been introduced in a
phenomenological way.

It has been shown in ref. \cite{Biondini:16a,Biondini:16b} that the
development of the oscillation behavior in the nonlinear stage of MI
does not depend on the exact shape of the localized perturbation,
provided some (relatively mild) conditions necessary for a rigorous
treatment are satisfied. We have investigated this point from our
experiments. Fig. 3 shows  space-time evolution of an optical plane
wave initially perturbed by a localized dark (negative) perturbation.
Experimental results plotted in the left part of Fig. 3 reveal that a
nonlinear oscillatory structure grows from the initial perturbation.
One can see that the detailed structure exhibiting two symmetric
solitary waves separated by a narrow ``vacuum'' region in the central
part slightly differs from the one observed in the positive
perturbation case. However, the leading order modulation solution
describing this structure is  the same as in the positive perturbation
case. This can be readily understood by noticing that the dynamics of
the plane wave under sufficiently negative localized perturbation can
be viewed as a combination of two focusing dam breaks of opposite
signs located close to each other. It follows from the results of
refs. \cite{Kamchatnov:97, GEl:16, Jenkins:14} that the modulation
solution describing such `double dam break' problem is exactly the
same as the one from refs. \cite{GEl:93,Biondini:16a,Biondini:16b,
  Biondini:17} for the positive perturbation case (see Appendix \ref{sec:MI})
, confirming the universality of the observed structure.

As shown in the right panel of Fig. 3, the behavior  observed
experimentally  is also quantitatively well described by the numerical
simulation of Eq. (\ref{eq:NLSE}) taking $\psi(z=0,t)=\sqrt{P_0 (1 -
  \beta \exp(-t/T_0)^4)}$ as the initial condition ($P_0=16$ mW,
$\beta=0.93$, $T_0=50$ ps) that fits the experimental profile shown in
Fig. 3(b). 

As shown in refs. \cite{GEl:93,Biondini:16a}, the boundaries of the
region separating the nonlinear oscillatory solution from the plane
wave region are expanding linearly with the evolution variable. When
rephrased in physical variables, these boundaries are given by $t_\pm
= \pm 2 \sqrt{2 \beta_2 \gamma P_0} z$ (see Appendix \ref{sec:MI}).
They are plotted with white straight lines  in Fig. 2(a)(d) and
Fig. 3(a)(d). Even though the oscillatory structure is effectively
located within the linear boundaries, the edges of the nonlinear
oscillating structure are relatively far from the  boundaries
predicted by the asymptotic (long-time) theory. This quantitative
difference between experiment and theory arises from the fact that the
theoretical result has been established in the framework of the purely
integrable and {\it non-dissipative} 1D-NLSE, see Appendix \ref{sec:damping}.
Further theoretical work
is needed to take into account the influence of a small linear damping
term on the mathematical expression giving the boundaries separating
the oscillatory region from the plane wave region.

In summary, we have reported an optical fiber experiment in which we
have observed the space-time dynamics of a modulationally-unstable
plane wave perturbed at initial time by a localized peak. Our
experimental results demonstrate the robustness to noise and
dissipation of the expanding modulated solution theoretically found in
ref.  \cite{GEl:93,Biondini:16a,Biondini:16b}. Our experimental
platform could be further used to explore some other scenarios of the
nonlinear stage of MI, including integrable turbulence or soliton gas
\cite{Zakharov:09,Randoux:17,GEl:05,Gelash:18}. 

\begin{acknowledgments}
This work has been partially supported  by the Agence Nationale de la
Recherche  through the LABEX CEMPI project (ANR-11-LABX-0007),   the
Ministry of Higher Education and Research, Hauts de France council and
European Regional Development  Fund (ERDF) through the the Nord-Pas de
Calais Regional Research Council and the European Regional Development
Fund (ERDF) through the Contrat de Projets Etat-R\'egion (CPER
Photonics for Society P4S). The work of GE was partially supported by
EPSRC grant EP/R00515X/1.  The authors are grateful to Draka-Prysmian
for fiber supplying and L. Bigot, R. Habert, E. Andresen and
IRCICA-TEKTRONIX European Optical and Wireless Innovation Laboratory
for technical support about the electronic devices.  The authors are
also grateful to A. Mussot, C. Naveau and P. Szriftgiser for providing
temporary access to some specific optical filter.   
\end{acknowledgments}

\appendix



\section{Details of the optical fiber experiment}\label{sec:exp}

In this section, we provide more details on the experimental
setup sketched in Fig. 1. 

The light source used for generation of the plane wave is
a single-frequency continuous-wave laser diode (APEX-AP3350A)
centered at $1550$ nm which
delivers an optical power of a few mW. The short localized perturbation
of the plane wave is first generated by means of an electro-optic
modulator (EOM, NIR-MX-LN series, bandwidth $20$ GHz, Photline)
driven by an arbitrary waveform generator (AWG70000, bandwidth $50$ GHz,
Tektronix). The arbitrary waveform generator generates
a periodic square electrical signal with a duty cycle
of $2 \times 10^{-4}$, i.e. the duration of the square
electrical pulses is $20$ ps and the period of the signal is
$100$ ns. At the output of the EOM, the light wave
has a constant (infinite) background periodically modulated
by pulses having a width of $30$ ps and a period
of $100$ ns. The bright (positive) or dark (negative) nature of
the peaks modulating the cw bakground can be selected
by changing the bias voltage applied to the EOM. The constrast of
the modulated train of pulses can be varied by changing the
control voltage applied to the EOM.

The power of the modulated
wave is amplified to the Watt-level by using an 
Erbium-doped fiber amplifier.
Then the amplified wave is optically
chopped by an acousto-optic modulator (AOM) that produces a periodic train
of square pulses having a width of $\sim 100$ ns and a
period of $10$ ms. With this method, we obtain 
a perturbed plane wave having a duration of $\sim 100$ ns which
is much greater than the typical duration ($ 30-50$ ps)
of the nonlinear structures that are observed. The optical power
launched inside the recirculating fiber loop can be adjusted by using
a half-wave plate placed before a polarizing cube inside an aerial arm
that is not shown in Fig. S1 for the sake of simplicity. 

The perturbated square pulse is monitored fast
photodiodes (Picometrix D-8IR) connected to a fast oscilloscope
(LeCroy Labmaster 10-65ZI) having
a bandwidth of $65$ GHz and a sampling rate of $80$ GSa/s. 
The photodiode PD1 has been carefully calibrated in some
annex experiment and the voltage measured can be 
converted into optical power with a relative accuracy
that is below $\sim 10 \%$.

The recirculating fiber loop is made up of
$\sim 4$ km of single-mode fiber (SMF)
closed on itself by a $90/10$ fiber coupler. The
coupler is arranged in such a way that $90 \%$ of the intracavity
power is recirculated. The SMF has been manufactured by Draka-Prysmian.
It has a measured second-order dispersion coefficient of
$-22$ ps$^2$ km$^{-1}$ and
an estimated Kerr coefficient of $1.3$ km$^{-1}$ W$^{-1}$
at the working wavelength of $1550$ nm. 

Raman amplification is achieved by injecting a pump  wave
$1450$ nm inside the fiber loop. The pump wave is coupled
in and out the recirculating fiber loop by using two
commercial wavelength dense multiplexers (WDMs) that split
light at $1450$ nm and at $1550$ nm into two separate fiber
paths. The distance between the two WDMs inside the recirculating
fiber loop is $\sim 2$ km. The pump laser at $1450$ nm is
a commercial Raman fiber laser delivering an output beam
having a power of several Watt. In our experiments,
this optical power is attenuated to typically $\sim 500$ mW
by using a $90/10$ fiber coupler (not shown in Fig. 1).

\section{Modulational instability of a locally perturbed plane wave and the focusing dam break problem}\label{sec:MI}

\begin{figure*}[!t]
  \includegraphics[width=0.8\textwidth]{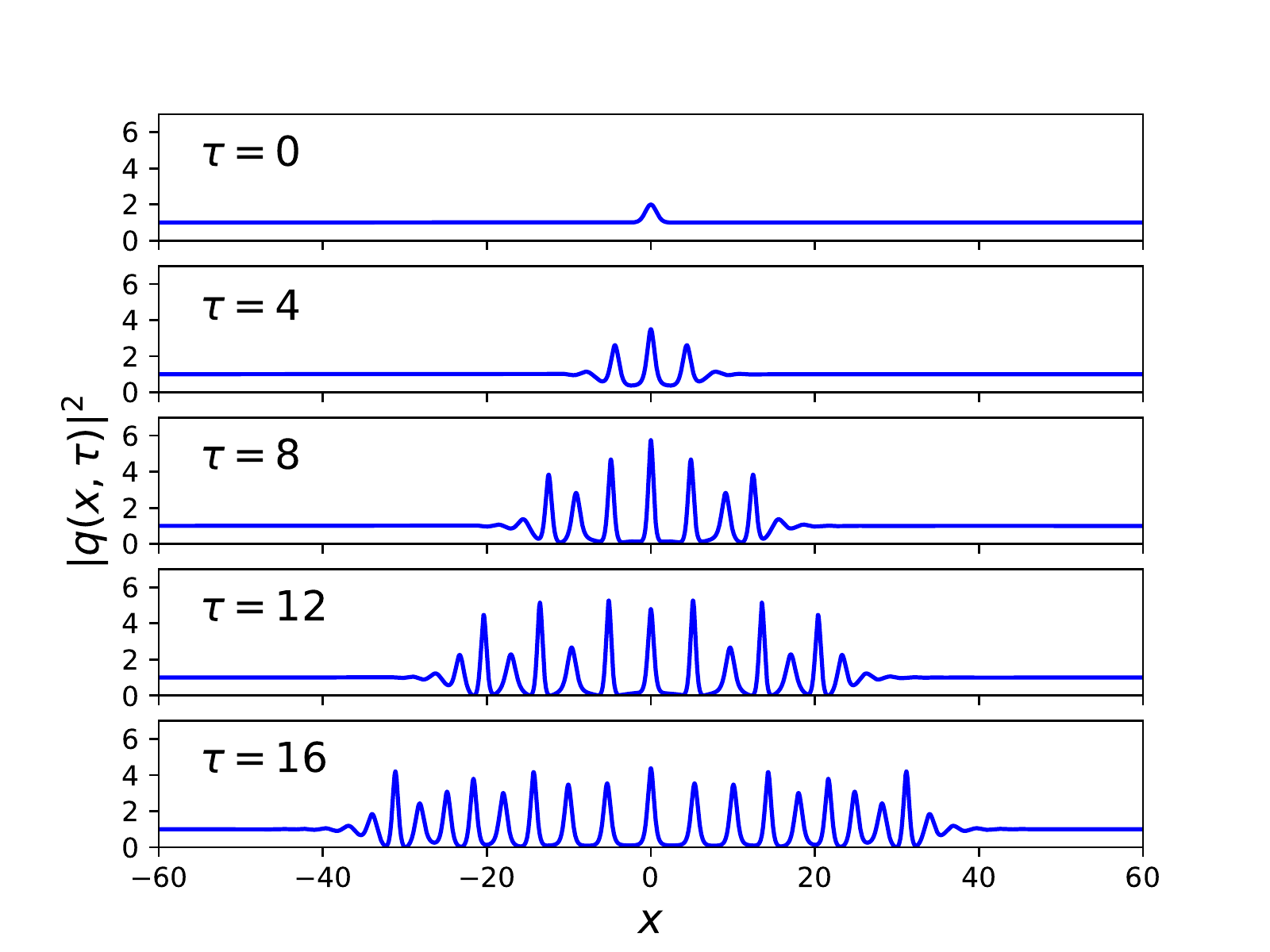}
  \caption{Evolution of a locally perturbed plane wave towards the universal modulated elliptic wave pattern. Numerical simulation of Eq. \eqref{nls1} with $\varepsilon=1$ and initial data \eqref{ic1} with $q_0=1$ and $\delta(x) =(\sqrt{2}-1) \exp(-x^2)$. The fiber optics experiment with the positive perturbation reported in Fig. 2 accurately captures the initial stage of the evolution (around $\tau=4$).
  }
  \label{figS1}
\end{figure*}

We consider the 1D-NLSE in the  form
\begin{equation}\label{nls1}
i  \varepsilon q_{\tau} +  \frac12 \varepsilon^2 q_{xx} + |q|^2 q = 0 \, ,
\end{equation}
where $\varepsilon$ is the normalized dispersion parameter. The relations between the variables in Eq.~\eqref{nls1} and  Eq.~(1)  are
\begin{equation}\label{relations}
q= \psi/\sqrt{P_0}, \quad \tau = z/\sqrt{L_{NL}L_D}, \quad x= t/T_P,  \quad \varepsilon =  \sqrt{L_{NL}/L_D},
\end{equation}
where $T_P$ represents the large initial scale, i.e. the duration of
the square pulses in the experiment ($\sim 100$ ns). $L_{NL}=1/(\gamma P_0)$ and
$L_D=T_P^2/|\beta_2|$ represent the nonlinear length and the linear dispersive
length, respectively.

We consider Eq.~\eqref{nls1} with the perturbed plane wave initial data 
\begin{equation}\label{ic1}
q(x,0)= q_0(1 + \delta(x)),
\end{equation}
where $q_0>0$ and $\delta(x)$ decays sufficiently rapidly as $|x| \to \infty$ (see \cite{Biondini:16a,Biondini:16b} for detailed restrictions on  $\delta(x)$.)

The {\it long-time} ($\tau \gg 1$) asymptotic solution of  \eqref{nls1}, \eqref{ic1} was found in \cite{Biondini:16a,Biondini:16b}  to be $q=Q(x,\tau) + O(1/\sqrt{\tau})$, where the leading term $Q(x,\tau)$ has different form in different regions of $x$-$\tau$ plane: $Q(x,\tau)=q_0$ for $x<-2\sqrt{2}q_0 \tau$ and $x>2\sqrt{2}q_0 \tau$ while for 
$ x \in [-2\sqrt{2}q_0 \tau,  2\sqrt{2}q_0 \tau]$ it has the form of slowly modulated travelling wave:
\begin{equation}\label{elliptic}
|Q(x,t)|^2=(q_0+b)^2 - 4q_0b \, \hbox{sn}^2 \left( 2\sqrt{q_0b/m} \, (x-a \tau - x_0)\varepsilon^{-1}; m \right),\end{equation}
where $\hbox{sn}(\cdot)$ is a Jacobi elliptic function with the modulus $ m \in [0,1]$ given by
$$
m=\frac{4q_0 b}{a^2+(q_0+ b)^2}\, .
$$
The modulation parameters $a(x, \tau)$, $b(x,\tau)$ are  found from equations
\begin{widetext}
\begin{equation}\label{mod}
\begin{split}
a= \sigma  \frac{2q}{m \mu(m)}\sqrt{(1-m)[\mu^2(m)+m -1]} ,  \qquad  b = \frac{q}{m \mu(m)}[(2-m)\mu(m) -2(1-m)] \, ,  \\
\frac{x}{\tau}= \sigma \frac{2q}{m \mu(m)}\sqrt{(1-m)(\mu^2(m)+m-1)} \left ( 1+ \frac{(2-m)\mu(m) -2(1-m)}{\mu^2(m) +m - 1}\right) \, ,
\end{split}
\end{equation}
\end{widetext}
where $\sigma =  \hbox{sgn}(x)$ and  $\mu(m)=E(m)/K(m)$, where $K(m)$ and  $E(m)$ are the complete elliptic integrals of the first and second kind respectively.  The initial phase $x_0$ in \eqref{elliptic} depends on the specific form of the perturbation function $\delta(x)$. The modulated elliptic solution \eqref{elliptic}, \eqref{mod} was originally obtained in \cite{GEl:93} in the framework of the  Whitham modulation theory \cite{whitham}, where the universal modulation \eqref{mod} was found as a  unique self-similar solution of the Whitham-NLSE equations satisfying the plane wave conditions at infinity. 

Solution \eqref{elliptic}, \eqref{mod} describes a symmetric expanding oscillatory structure having the form of the fundamental soliton ($m=1$) resting at $x=0$, and degenerating, via the modulated elliptic regime, into a vanishing amplitude linear wave ($m=0$) at the edges propagating away with the velocities $\pm 2 \sqrt{2} q_0$. Importantly, modulation solution \eqref{mod} depends only on the plane wave background $q_0$ and does not depend on the shape or amplitude of the initial perturbation $\delta(x)$ and is, therefore, universal \cite{Biondini:16a,Biondini:16b}. 

Using the transformations \eqref{relations} we find the  location of the expansion wedge in the  physical $t$-$z$ plane: $t_{\pm} = \pm 2 \sqrt{2 \beta_2 \gamma P_0} z$.

The evolution of a localized initial perturbation towards the universal asymptotic pattern described by \eqref{elliptic}, \eqref{mod} is shown in Fig.~\ref{figS1}. In the experiment reported in Fig.~2 the initial stage of the evolution is robustly captured.

Remarkably, solution \eqref{elliptic}, \eqref{mod} considered for $x<0$ ($x>0$) and complemented by $q=0$ for $x>0$ ($x<0$) also describes (up to the initial phase $x_0$) the leading order term in the  asymptotic solution of the 1D-NLSE {\it focusing dam break problem} \cite{Kamchatnov:97, GEl:16, Jenkins:14}
\begin{equation}\label{ic2}
q(x,0)  = q_\pm(x) =\left\{
\begin{array}{ll}
q_0  &\quad \hbox{for} \quad \pm x <0,\\
 0& \quad \hbox{for} \quad \pm x >0 \, .
\end{array}
\right.
\end{equation} 
(the `$\pm$' refers to the two possible dam break configurations with `step down' ($+$) and `step up' ($-$))
The long time asymptotic solutions of \eqref{nls1}, \eqref{ic2} are then given by $q(x,t)=Q_{\pm}(x, \tau) + O(1/\sqrt{\tau})$ with
\begin{equation}\label{Qpm}
Q_\pm(x, \tau) =\left\{
\begin{array}{ll}
Q(x, \tau)  &\quad \hbox{for} \quad \pm x <0,\\
 0& \quad \hbox{for} \quad \pm x >0 \, .
\end{array}
\right.
\end{equation} 
and the modulation  \eqref{mod} with the appropriate $\sigma$ chosen.  
We note that the rigorous analysis of ref.~\cite{Jenkins:14} was performed for  {\it decaying} initial conditions in the form of a rectangular barrier of the width $L$ so the solution \eqref{Qpm}, \eqref{elliptic}, \eqref{mod} is rigorously valid as the leading term of the  {\it small-dispersion} ($\varepsilon \ll 1$) asymptotic $q= Q(x,\tau)+ O(\varepsilon^{1/2})$ for $0<\tau < L/4\sqrt{2}q_0$. 

One can see now that the leading term of the asymptotic solution for the locally perturbed plane wave is equivalent to the leading term of the solution to the `double dam break' problem, 
$$
Q(x, \tau) = Q_-(x, \tau) + Q_+(x, \tau).
$$
We note that within the Whitham modulation theory the validity of  essentially the same solution  \eqref{elliptic}, \eqref{mod} for both the locally perturbed plane wave and the dam break data is to due self-similarity of both problems on the level of modulation equations. The modulation self-similarity argument also explains persistence of the same universal pattern  reported in \cite{Biondini:17} beyond the integrable 1D-NLSE. Indeed, the Whitham modulation theory does not have integrability of the original dispersive equation as a pre-requisite hence self-similar modulation solutions are in principle available for non-integrable systems \cite{GEl1:05}.

\begin{figure*}[!t]
  \includegraphics[width=0.8\textwidth]{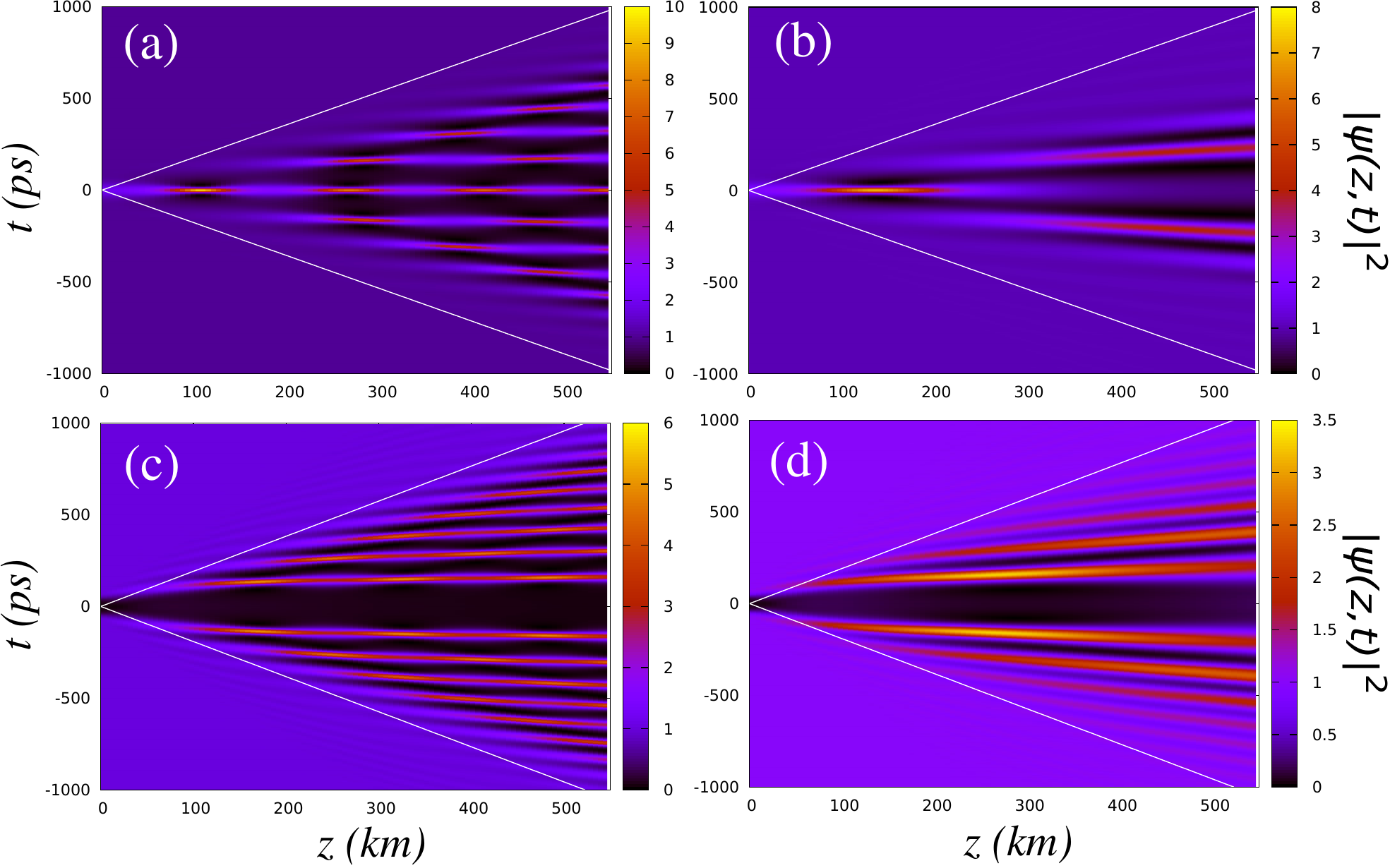}
  \caption{Numerical simulations 
    Eq. (1) showing the influence of a
    small linear damping term on the development
    of the nonlinear oscillatory structure while starting from
    (a), (b) a bright (positive) perturbation and (c), (d)
    a dark (negative) perturbation. The numerical simulations 
    are made with the following parameters:
    $\beta_2=-22$ ps$^2$km$^{-1}$, $\gamma=1.3$W$^{-1}$km$^{-1}$.
    In (a) and (c), $\alpha_{eff}=0$ whereas in
    (b) $\alpha_{eff}=4.2 \times 10^{-3}$ km$^{-1}$
    and in (d) $\alpha_{eff}=4 \times 10^{-3}$ km$^{-1}$.
    In (a), (b), the initial contion is
    $\psi(z=0,t)=\sqrt{P_0 (1 + \exp(-t/T_0)^2)}$) with $P_0=14$ mW, $T_0=30$ps.
    In (c), (d), the initial condition is
    $\psi(z=0,t)=\sqrt{P_0 (1 - \beta \exp(-t/T_0)^4)}$
    with $P_0=16$ mW, $T_0=50$ ps, $\beta=0.93$. 
  }
  \label{figS3}
\end{figure*}

\section{Influence of linear damping on the space-time evolution
of the localized perturbations}\label{sec:damping}

In this section, we use numerical simulations of Eq. (1) to
give clear evidence of the role of linear damping (unavoidable in the
experiment) on the development of the nonlinear oscillatory structure that
grows from the initial localized pertubation.  If the loss term
$\alpha_{eff}$ in Eq. (1) is set to zero, Fig. \ref{figS3}(a)
and \ref{figS3}(c) show that the nonlinear oscillatory structure
expands linearly with the propagation distance inside the fiber.
The edges of the nonlinear oscillating structure coincide
well with the boundaries predicted by the asymptotic
(long-time) theory, in full agreement with theoretical
results reported in ref. \cite{Biondini:16a,Biondini:16b,GEl:93}.
These boundaries are given
by $t_{\pm} = \pm 2 \sqrt{2 \beta_2 \gamma P_0} z$ (see Appendix \ref{sec:MI})
and they are plotted with white lines in Fig. \ref{figS3}.

The expansion of the nonlinear structure is strongly
influenced by the presence of small linear dissipation, as
shown in space-time diagrams of Fig. \ref{figS3}(b) and
\ref{figS3}(d) that have been computed with $\alpha_{\rm eff} \ne 0$. .
Even though the oscillatory structure is effectively located
within the linear boundaries (white lines in Fig. \ref{figS3}(b) and
\ref{figS3}(d)), the
edges of the nonlinear oscillating structure are relatively
far from the boundaries predicted by the asymptotic
(long-time) theory, in particular for the initial condition
where the localized perturbation is bright at initial
time (Fig. \ref{figS3}(b)). The importance of
weak linear fiber losses has already been pointed out in the
experiments of ref. \cite{Xu:17} where the 
dispersive dam-break flow of a photon fluid (defocusing regime) has been
observed. Further theoretical work is needed to take into account the
influence of a small linear damping term on the mathematical
expression giving the boundaries separating the
oscillatory region from the plane wave region.




\end{document}